\documentclass[preprint,amsmath,amssymb, aps, prl, floatfix]{revtex4-1}
\usepackage{graphicx}
\usepackage{dcolumn}
\usepackage{bm}
\usepackage{caption}
\usepackage{amssymb,amsmath,color}
\usepackage{gensymb}
\usepackage{url}
\usepackage{float}
\usepackage{hyperref}    
\hypersetup{colorlinks = true, linkcolor = blue, anchorcolor = blue, citecolor = blue, filecolor = blue, urlcolor = blue}

\begin{document}

\title{Reconstruction of moir\'{e} lattices in twisted
transition metal dichalcogenide bilayers}

\author{Indrajit Maity, Prabal K. Maiti,
H.~R.~Krishnamurthy} \author{ Manish Jain}
\email{mjain@iisc.ac.in}
\affiliation{ Centre for Condensed Matter Theory, Department
of Physics, \\ Indian Institute of Science, Bangalore-560012
}%

\begin{abstract}
An important step in understanding the exotic electronic,
vibrational, and optical properties of the moir\'{e} lattices is
the inclusion of the effects of structural relaxation of the
un-relaxed moir\'{e} lattices. Here, we propose novel structures
for twisted bilayer of transition metal dichalcogenides (TMDs). For
$\theta\gtrsim 58.4^{\circ}$, we show a dramatic reconstruction of
the moir\'{e} lattices, leading to a trimerization of the
unfavorable stackings. We show that the development of curved
domain walls due to the three-fold symmetry of the stacking energy
landscape is responsible for such lattice reconstruction.
Furthermore, we show that the lattice reconstruction notably
changes the electronic band-structure. This includes the occurrence
of flat bands near the edges of the conduction as well as valence
bands, with the valence band maximum, in particular, corresponding
to localized states enclosed by the trimer. We also find
possibilities for other complicated, entropy stabilized, lattice
reconstructed structures. \end{abstract}

\maketitle 

The formation of flat bands in the electronic band structure
of moir\'{e} patterns of two-dimensional materials is
central to understanding the observed exotic electronic
phases \cite{bistritzer2011moire,
cao2018correlated,cao2018unconventional,
yankowitz2019tuning}. Twisted bilayer TMDs can possess flat
bands for a continuum of twist angles
\cite{naik2018ultraflatbands,naik2019origin,fleischmann2019moir,wang2019magic,zhang2019flat,
wu2019topological,zhen2019designing,angeli2020gamma,pan2020band,zhan2020tunability,zhai2020theory}.
To accurately calculate their electronic band structure,
incorporation of structural relaxation effects is crucial
\cite{naik2019origin,gargiulo2017structural,yoo2019atomic,naik2018ultraflatbands,
lucignano2019crucial, nam2017lattice, leconte2019relaxation,
halbertal2021moire}.  Typically, these relaxations are
performed by starting from a configuration and \textit{only}
allowing downhill motion in the potential energy landscape
using local search algorithms (standard minimization). Since
the number of local minima in the potential energy landscape
increases exponentially with the number of atoms, standard
minimizations are often insufficient for finding the stable
structures\cite{pickard2011ab, stillinger1999exponential,
kirkpatrick1983optimization}. All the studies conducted on
moir\'{e} materials to date presume that the moir\'{e}
lattice constant of the un-relaxed twisted structure remains
intact even after relaxation.
\cite{bistritzer2011moire,cao2018correlated,cao2018unconventional,yankowitz2019tuning,naik2018ultraflatbands,naik2019origin,fleischmann2019moir,wang2019magic,zhang2019flat,wu2019topological,gargiulo2017structural,yoo2019atomic,lucignano2019crucial,nam2017lattice,carr2018relaxation,enaldiev2019stacking,weston2019atomic,rosen2020twist,leconte2019relaxation}. 

Here, from the structures obtained using simulated annealing
we demonstrate that a dramatic reconstruction of moir\'{e}
lattices of TMDs takes place for $\theta \gtrsim
58.4^{\circ}$. Thus, the presumption that the moir\'{e}
lattice constant of the rigidly twisted structures continues
to characterize the relaxed structures is not always valid.
Such lattice reconstructions are not accessible in standard
minimization approaches. We discuss below the details of the
lattice reconstruction for twisted bilayer (tBL) of
$\mathrm{MoS_{2}}$. We have also verified our conclusions
for $\mathrm{MoSe_{2}, \ WSe_{2}, \ WS_{2}}$ (see
Supplementary Information (SI), Sec.~II \cite{SI}). We
demonstrate that the lattice reconstruction substantially
changes the electronic band structure.  

We use the TWISTER code \cite{naik2018ultraflatbands} to
construct tBLTMDs. We use the Stillinger-Weber and
Kolmogorov-Crespi (KC) potential to capture the intra and
interlayer interaction of tBLTMDs, respectively
\cite{SW_arxiv_2017, Mit_kc_2018}. The used KC parameters
have been shown to accurately capture the interlayer van der
Waals interaction present in the TMDs \cite{Mit_kc_2018}. We
relax the tBLTMDs in LAMMPS using standard minimization
\cite{Plimpton_jcp_1995,bitzek2006structural}, denoted as
standard relaxation (SR). We also perform classical
molecular dynamics simulations using the canonical ensemble
at $T=1$ K and cool down snapshots to 0 K, and then carry
out an energy minimization. We refer to this second approach
as simulated annealing (SA). The phonon frequencies are
calculated using modified PHONOPY\cite{phonopy} code. We
perform electronic structure calculations using density
functional theory\cite{kohn1965self} with SIESTA
\cite{lin2014siesta, soler2002siesta, yu2018elsi,
lin2009fast,troullier1991efficient,dion2004van,cooper2010van,van2018pseudo} (SI Sec.~I for details
\cite{SI}).

\begin{figure*}[!htbp]
\includegraphics[width=\textwidth]{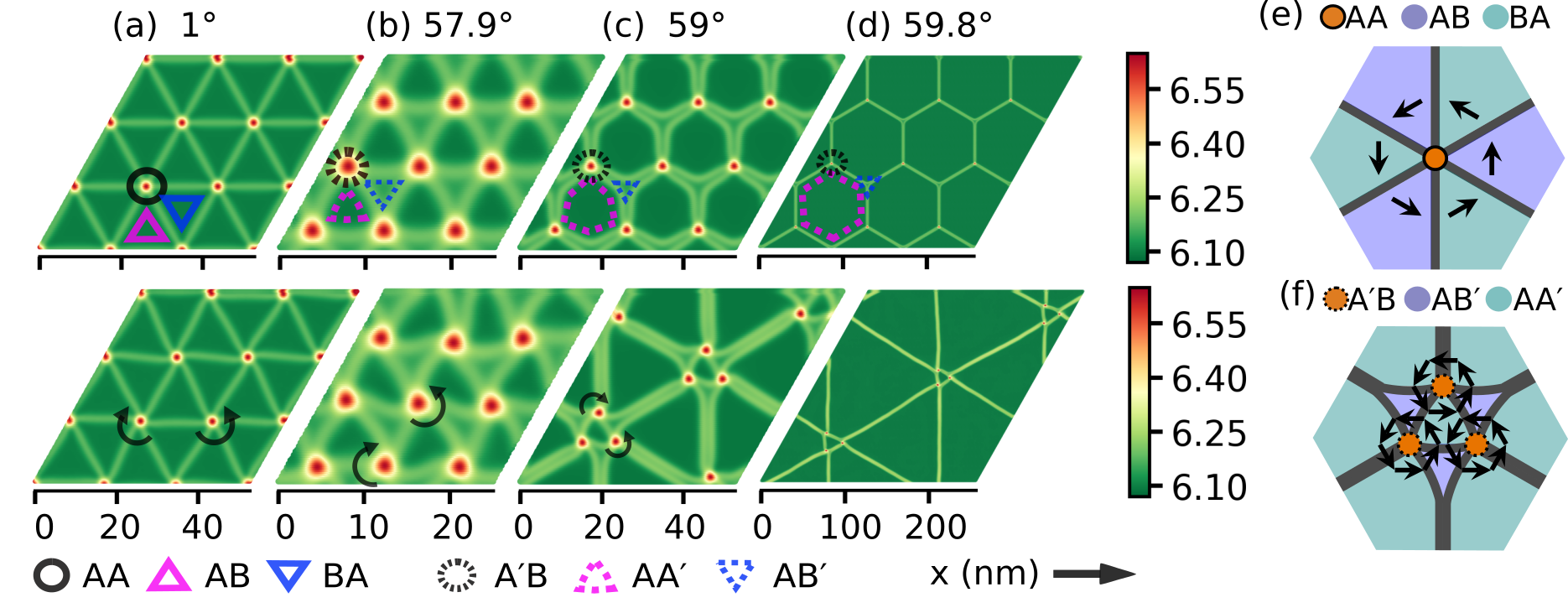}
\caption{(a)-(d): Interlayer separation landscape of
$\mathrm{tBLMoS_{2}}$ using standard relaxation (top panel)
and simulated annealing (bottom panel). The smallest
repetitive cell in the top panel is a moir\'{e} unit cell.
The scales of the colorbar, in \AA\ and corresponds to
interlayer separation. The curling of domain walls near a
few $\mathrm{AA,\ A^\prime B}$ stackings are marked.
(e),(f): Schematics near the topological defects for
$\mathrm{tBLTMDs}$ for $\theta=1^{\circ},\ 59^{\circ}$,
respectively. The order parameter is shown with arrows.}
\end{figure*}

Due to the presence of different sub-lattice atoms (Mo/W,
S/Se) in TMD, the tBLTMD possesses distinct high-symmetry
stackings for $\theta$ near $0^{\circ}$ ($\mathrm{AA,\ AB,\
BA}$) and near $60^{\circ}$ ($\mathrm{AA^\prime \ ,AB^\prime
, \ A^\prime B  }$) \cite{liang2017interlayer}. Nevertheless,
the lattice constants of the un-relaxed tBL are identical
for $\theta$ and $60^{\circ}- \theta$ (e.g. $1^{\circ}$ and
$59^{\circ}$). Among the above mentioned stackings,
$\mathrm{AB}$ is energetically the most favourable stacking
as $\theta \to 0^{\circ}$ ($E_{\mathrm{AB}} =
E_{\mathrm{BA}} < E_{\mathrm{AA}}$, six-fold symmetric
around $\mathrm{AA}$) and $\mathrm{AA^\prime}$ for $\theta
\to 60^{\circ}$ ($E_{\mathrm{AA^\prime}} <
E_{\mathrm{AB^\prime}} < E_{\mathrm{A^\prime B}}$,
three-fold symmetric around $\mathrm{A^\prime B}$)
\cite{Mit_kc_2018,carr2018relaxation}.

In Fig.~1 we show the interlayer separation (ILS) landscape
for a $3\times 3\times 1$ moir\'{e} supercell of
$\mathrm{tBLMoS_{2}}$, obtained using both SR and SA. The
landscape for $\theta=1^{\circ}$ is a representative of
$\theta\to0^{\circ}$ (Fig.~1a, top panel). With SR we find
straight domain walls separating AB, BA stackings. On the
other hand, the ILS landscape computed with SA shows a
slight curling of the domain walls near AA stacking
(Fig.~1a, bottom panel). Although the
number of clockwise and counter-clockwise curlings are equal,
they do not always form a checkerboard-like pattern. While
the checkerboard pattern is the lowest in energy, the
energy difference between the checkerboard pattern and a
random distribution of curlings is small (a few meV per
moir\'{e} lattice). Nevertheless, the AA stackings always
form a triangular lattice for any $\theta$ close to
$0^{\circ}$, consistent with experiments
\cite{weston2019atomic, rosen2020twist, zhang2019flat}. 

In contrast, the behavior of the ILS landscape shows very
different, and intriguing features as $\theta\to60^{\circ}$.
We categorize the $\theta$ dependence into two regions.
Region I ($\theta <58.3^{\circ}$) : With SR both the
$\mathrm{AA^\prime}$ and $\mathrm{AB^\prime}$ stackings
occupy comparable areas of the supercell, with each forming
an approximate equilateral triangle (Fig.~1b). Similar to
$\theta \to 0^{\circ}$, the ILS landscape obtained with SA
shows curlings of domain walls near $\mathrm{A^\prime B}$
stacking (Fig.~1b, bottom panel). Region II ($\theta \gtrsim
58.4^{\circ} $) : The most favorable ($\mathrm{AA^\prime}$)
stacking increases in area significantly and evolves from
Reuleaux triangles to approximate hexagonal structures, as
obtained with SR (Fig.~1c-d, top panel), consistent with
previous studies \cite{carr2018relaxation,
enaldiev2019stacking}. In this case, the domain walls
connecting $\mathrm{A^\prime B}$ stackings are significantly
curved and never straight-lines. These latter structures \textit{show
notable reconstruction with SA}. In particular, a triangular
lattice is formed with three $\mathrm{A^\prime B}$ stackings
trimerizing to form a motif (Fig.~1c-d, bottom panel).
Moreover, the domain walls connecting different
$\mathrm{A^\prime B}$ stackings are almost straight in the
reconstructed structures. The reconstructed structures obtained
using SA are always energetically more stable than those
obtained using SR. 

We characterize the domain walls using the order-parameter,
defined as the shortest displacement vector required to take any
stacking to the most unfavorable stacking
\cite{alden2013strain,naik2018ultraflatbands,gargiulo2017structural}.
Irrespective of $\theta$, we find the domain walls to be
shear solitons (change in order parameter is along the
domain wall as we go from $\mathrm{AB}\to\mathrm{BA}$ for
$\theta\to0^{\circ}$ and
$\mathrm{AA^\prime}\to\mathrm{AA^\prime}$ for
$\theta\to60^{\circ}$). In Region II, two domain walls come
close together and the effective width increases. For
$\theta\to0^{\circ}$ ($\theta\to60^{\circ}$), the calculated
widths of the domain walls are : 2.9 (4.3) for
$\mathrm{tBLMoS_{2}}$, 2.9 (3.8) for $\mathrm{tBLMoSe_{2}}$,
3.7 (4.7) for $\mathrm{tBLWSe_{2}}$, 3.5 (4.5 ) for
$\mathrm{tBLWS_{2}}$ ( all in nm). Our estimated domain wall
widths are in good agreement with experiment
\cite{weston2019atomic}. Moreover, the order parameter
rotates by $2\pi$ at $\mathrm{AA/A^\prime B}$, indicating
it's topological nature (Fig.~1e, 1f). We do not find any
new creation or annihilation of the topological defects and
domain walls in our simulations.

\begin{figure}
\centering
\includegraphics[width=0.5\textwidth]{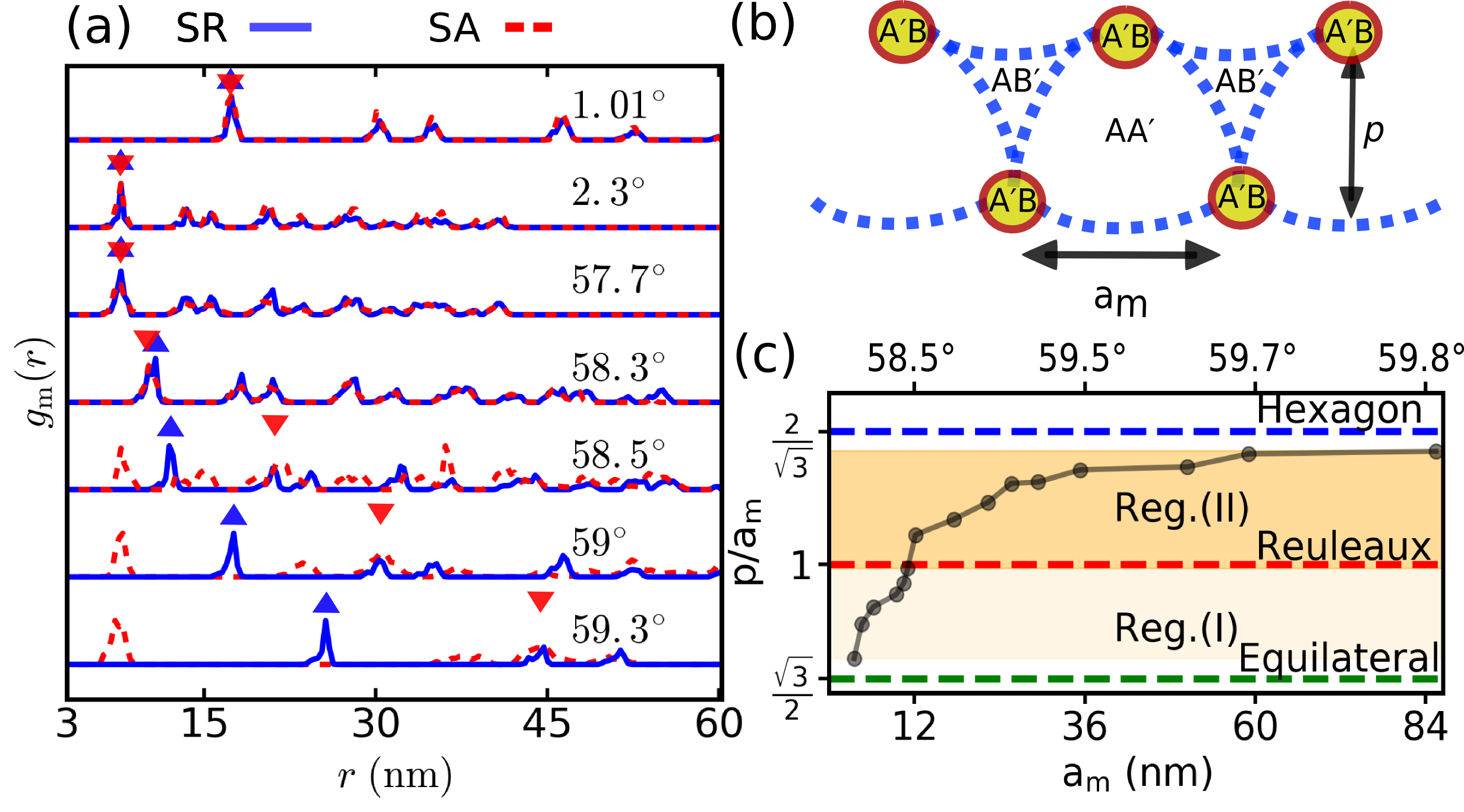}
\caption{(a) Radial distribution function,
$g_{\mathrm{m}}(r)$ computed with standard relaxation (SR)
and simulated annealing (SA). The moir\'{e} lattice
constants are marked (blue triangle pointing up for SR, red
triangle pointing down for SA). (b) Schematics of
$\mathrm{tBLMoS_{2}}$ as $\theta\to60^{\circ}$ with SR. (c)
Change of $p/\mathrm{a_{m}}$ with $\mathrm{a_{m}}$. Several
ideal geometric structures are marked with dashed lines.}
\end{figure}

\begin{figure*}[!htbp]
\includegraphics[width=\textwidth]{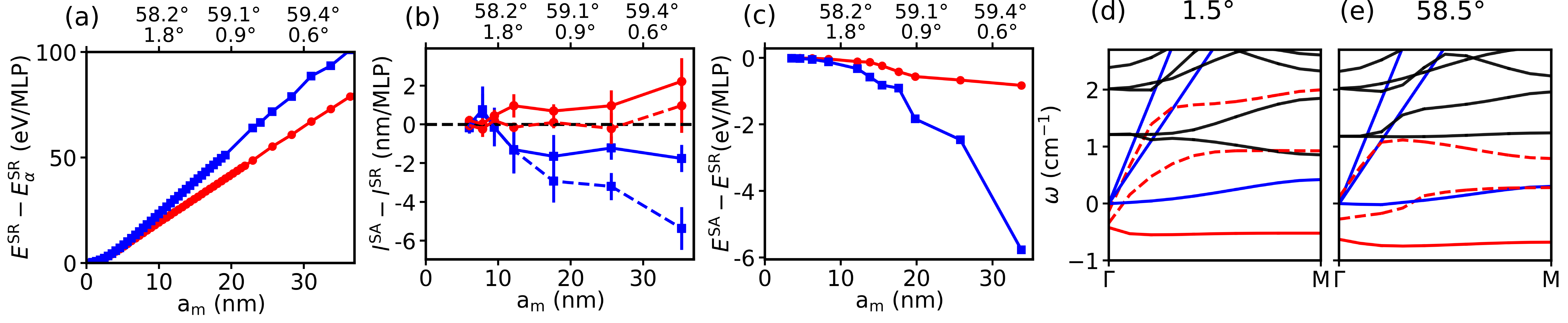}
\caption{(a) Change of total energy computed with respect to
stable stacking, $\alpha$ using SR with
$\mathrm{a_{m}}$ (corresponding $\theta$ are marked; blue
(red) for near $60^{\circ}$ ($0^{\circ}$) for (a)-(c)). (b)
Change in domain wall length with SA compared to SR
excluding (dashed lines) and including (solid lines) curling
near $\mathrm{AA/A^\prime B}$. The error bar denotes
standard deviation of the estimated change. (d) Total energy
gain with SA compared to SR. (d),(e) Phonon dispersion with
SR for $1\times 1\times 1$ moir\'{e} cell. The solid
blue, dashed red, solid red lines represent the acoustic,
phason, buckling mode localized at $\mathrm{AA/ A^\prime
B}$, respectively.} \end{figure*}

We investigate the structural long-range order by computing
the radial distribution function. In the $\mathrm{tBL}$
there are two distinct
length scales, one for the individual TMD layer given by the
lattice constant $a$, and the other for the moir\'{e}
lattice given by the $\theta$ dependent moir\'{e} lattice constant,
$\mathrm{a_{m}}=a/(2\sin (\theta /2))$. Therefore, we define
two separate radial distribution functions, one for atoms of
the individual layers and another for stackings of the
moir\'{e} lattice. We compute the moir\'{e}-scale radial
distribution function, $g_{\mathrm{m}}(r)$ using the
$\mathrm{AA/A^\prime B}$ stackings of $\mathrm{tBLMoS_{2}}$
(Fig.~2a). Each $\mathrm{AA/ A^\prime B}$ stacking
represents a moir\'e lattice point (MLP).

For $\theta \to 0^{\circ}$, $g_{\mathrm{m}}(r)$ obtained using SR and SA are similar (Fig.~2a). The average number of nearest
neighbour MLPs is always 6, calculated by integrating the
first peak of $g_{\mathrm{m}}(r)$. This confirms the
existence of the hexagonal network formed by domain walls
(Fig.~1a). Furthermore, the moir\'{e} lattice constant
calculated from $g_{\mathrm{m}}(r)$ is identical to that of
un-relaxed $\mathrm{tBLMoS_{2}}$. Therefore, the long-range
order of the un-relaxed structures remains intact as
$\theta\to0^{\circ}$. As $\theta\to60^{\circ}$ within Region I,
the moir\'{e} lattice constants are again identical for un-relaxed
and relaxed structures, $\mathrm{a_{m}
=a_{m}^{SR}=a_{m}^{SA}}$ (Fig.~2a), and the number of
nearest neighbour MLP is always 6. In contrast, the lattice reconstruction in Region II leads to the formation of a triangular
lattice with a modified lattice constant,
$\mathrm{a_{m}^{SA}=\sqrt{3}a_{m}}$ (Fig.~2a).
The first peak in the $g_{\mathrm{m}}(r)$ ($\approx 7.5$ nm)
corresponds to the motif of the triangular lattice. The motif
consists of 3 $\mathrm{A^\prime B}$ stackings. We find that
the number of the nearest neighbor of $\mathrm{A^\prime B}$
is 2. We also examine the atomic radial distribution
function for individual $\mathrm{MoS_{2}}$ layers.
Irrespective of $\theta$, the long-range order is preserved
at the unit-cell $\mathrm{MoS_{2}}$ scale. This establishes
that the aforementioned reconstruction in Region II is an
emergent phenomenon arising at the moir\'{e}-scale.

To pinpoint the onset of the lattice reconstruction
geometrically, we consider the ratio of the perpendicular
bisector, $p$, to $\mathrm{a_{m}}$ of $\mathrm{tBLMoS_2}$
obtained using SR (Fig.~2b). Interestingly, we find lattice reconstruction as $p/
a_{\mathrm{m}}$ becomes $\gtrsim 1$ (Fig.~2c). When
$p/\mathrm{a_{m}}=1$ ($\theta \sim 58.5^{\circ}$), the
$\mathrm{AA^\prime}$ stacking represents a Reuleaux triangle
with the domain walls occupying it's perimeter. When one
considers the perimeter$^{2}$ to area ratio, the Reuleaux
triangle is a local maximum \cite{modes2013spherical}. Since
the domain walls are energetically unfavorable compared to
$\mathrm{AA^\prime}$, the Reuleaux triangle is
expected to undergo rearrangements to minimize the total
energy. The shortest distance between two $\mathrm{A^\prime
B}$ stackings is $\approx d^{\mathrm{A^\prime B}} +
d^{\mathrm{AB^\prime}} \approx 3.3 + 4.5 = 7.8$ nm, where
$d^{\mathrm{A^\prime B}}, d^{\mathrm{AB^\prime}}$ denote the
sizes of the corresponding stackings. This explains the
occurrence of first peak in $g_{\mathrm{m}}(r)$ in Region II
at $\approx 7.5$ nm.

Next, we investigate the origin of these reconstructions
from energetics. The total energy of the
$\mathrm{tBLMoS_{2}}$ is a sum of the intralayer
energy, which is a combination of strain and bending energy
\cite{maity2018temperature}, and the interlayer energy. For
$\theta\to60^{\circ}$, the interlayer energy per MLP can be
approximated as, 
\begin{equation}
E_{\mathrm{inter}} - E_{\mathrm{inter}}^{\mathrm{AA^\prime}}
= \delta E^{\mathrm{A^\prime B}}_{\mathrm{inter}}
S^{\mathrm{A^\prime B}} + \delta
E^{\mathrm{DW}}_{\mathrm{inter}} S^{\mathrm{DW}} + \delta
E^{\mathrm{AB^\prime}}_{\mathrm{inter}}
S^{\mathrm{AB^\prime}} \end{equation}

Here, $\delta E^{\alpha}_{\mathrm{inter}}$ represents the
interlayer energy of stacking $\alpha$, evaluated with
respect to $\mathrm{AA^\prime}$ and $S^{\alpha}$ denotes the
occupied area. For small $\theta$, $S^{\mathrm{A^\prime B}},
\ S^{\mathrm{AB^\prime}}$ and the width of the domain wall
(DW), $w$, become constant ($S^{\mathrm{DW}}=wl$).
Therefore, the interlayer energy as in Eqn.(1) becomes
linear with the domain wall length, $l$, and is repulsive.
Moreover, the intralayer strain energies are concentrated on
the domain walls and scales as $l/w $
\cite{alden2013strain,zhang2018structural}. Thus, the
minimization of $l$ will minimize both the interlayer and
intralayer energies. In Fig.~3a we show the scaling of the
total energy with $\mathrm{a_{m}}$ using SR. The domain
walls obtained with SR are always significantly curved for
$\theta > 58.4^{\circ}$. The lengths of these curved domain
walls can be minimized by lattice reconstruction such that
the domain walls become straightlines (as in Fig.~1c,1d, bottom
panel). On the other hand, the domain walls are
straightlines for the corresponding set of $\theta$ near $0^{\circ}$
with SR. Thus, $l$ per moir\'{e} lattice is already
minimized. As a result, we do not find lattice
reconstruction with SA as $\theta\to 0^{\circ}$. However,
the domain walls obtained with SA are always curled near the $\mathrm{AA,\
A^\prime B}$ stackings, irrespective of lattice
reconstruction. This originates from a buckling instability,
primarily localized at $\mathrm{AA, A^\prime B}$ (see
below). Taking these into account, $l^{\mathrm{SA}}$ is
expected to be greater than $l^{\mathrm{SR}}$ in the absence
of lattice reconstruction. In Fig.~3b we show the estimate
of $(l^{\mathrm{SA}}-l^{\mathrm{SR}})$ per MLP as
$\theta\to0^{\circ},\ \to60^{\circ}$ (see SI, Sec.~III for
details). For $\theta \gtrsim 58.4^{\circ}$, the difference
becomes negative, indicating a reduction in the domain wall
length for the reconstructed lattice. The reduction in $l$,
disregarding the curling of the domain walls with SA, is large in Region II (Fig.~3b). Fig.~3c shows the gain in total
energy with SA relative to SR, which is significantly
greater in Region II than that for a corresponding $\theta$
near $0^{\circ}$ . The energy gain near $0^{\circ}$ arises
from curling of the domain walls near $\mathrm{AA}$, whereas
the gain in Region II arises predominantly from lattice
reconstruction.

We also compare the low-frequency vibrational modes of
$1.5^{\circ}$ and $58.5^{\circ}$ tBL$\mathrm{MoS_2}$. One of
the phason modes \cite{maity2019low} softens significantly
and becomes nearly dispersion-less with attributes of a
\textit{zero} mode for $58.5^{\circ}$ (Fig.~3d, 3e). Such a
mode is expected to cause reconstruction of lattices
\cite{sun2012surface}. Furthermore, with SR we find a soft
mode with imaginary frequency for both $1.5^{\circ}$ and
$58.5^{\circ}$ (Fig.~3d, 3e). The corresponding eigenvector
at $\Gamma$, which is localized on $\mathrm{AA/A^\prime B}$,
denotes a buckling instability and can be removed without
lattice reconstruction.  

\begin{figure}[!htbp]
\includegraphics[width=0.5\textwidth]{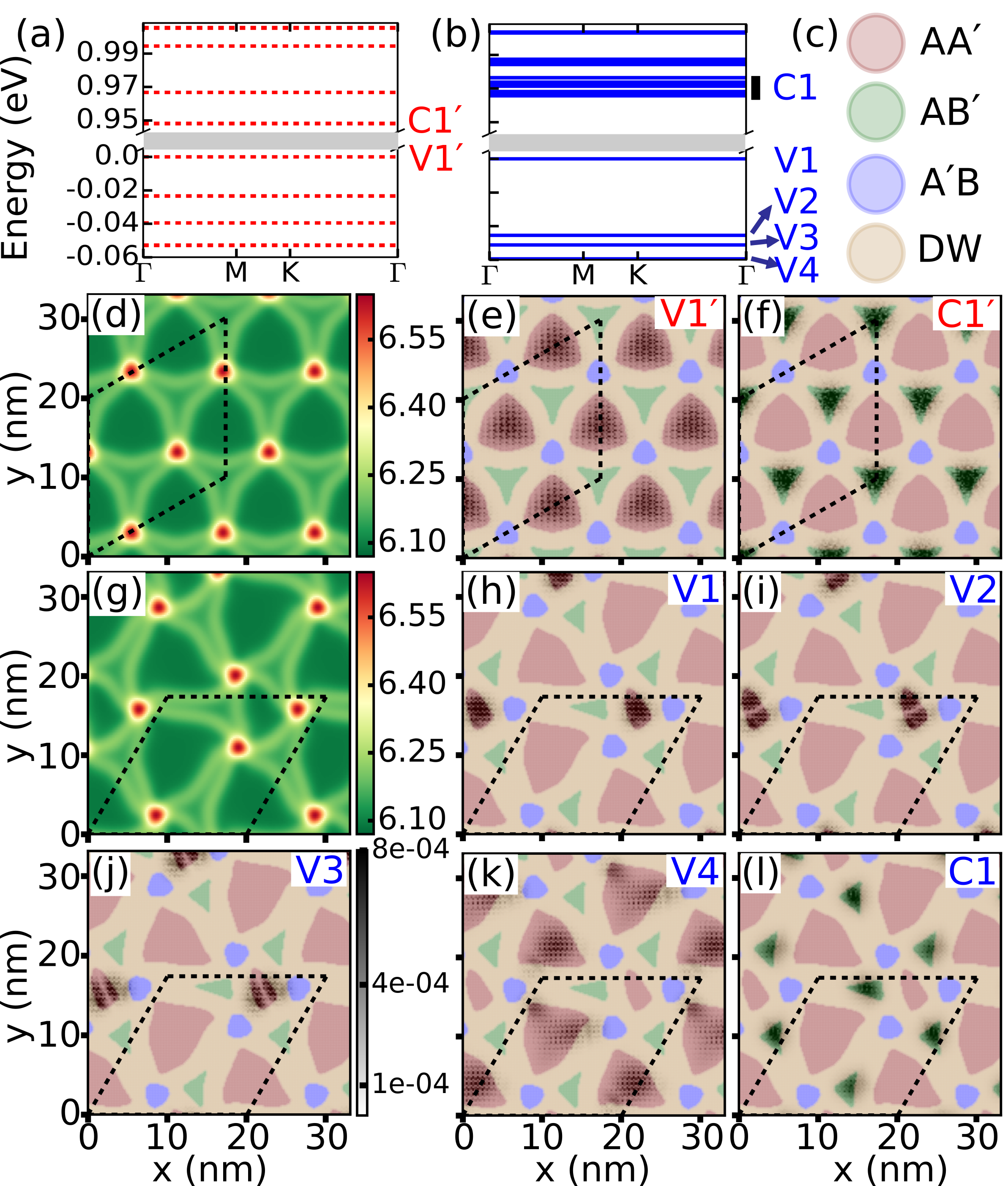}
\caption{(a),(b) Electronic band structures near the band edges of
a $\sqrt{3}\times\sqrt{3}\times1$ supercell of
$\mathrm{tBLMoS_{2}}$ for $58.47^{\circ}$ with SR and SA,
respectively. The supercell is marked with black dashed lines. (c)
Colors used to denote stackings in (e)-(f),(h)-(l). (d),(g) ILS
landscape for $58.47^{\circ}$ with SR and SA, respectively.
$|\psi_{\Gamma}(\vec{r})|^2$ averaged in the out-of-plane direction
of the states near VBM, and near CBM for structures obtained with
SR ((e)-(f)) and with SA ((h)-(l)) with the corresponding colorbar
shown in (j). A linear combination is shown in (l) as C1
corresponding to bands marked in (b).} \end{figure}

During our simulations we find transient structures, such as,
distorted hexagon, kagome, etc, which evolve to form the structures
shown in Fig~1c,d (SI, Sec.~IV). To investigate entropic effects,
we also simulate a supercell with 100 moir\'{e} lattices allowing
significantly large degrees of freedom for lattice reconstruction.
We find that lattice reconstructed structures with motifs of $>3$
$\mathrm{A^\prime B}$ stackings, nonuniform hexagons with parallel
domain walls are also possible (SI, Sec.~IV).  These structures can
be metastable due to the presence of a substrate, strain, etc in an
experiment. These external effects can modify the characteristic
angle for the onset of lattice reconstruction. Our study suggests
that the highly non-uniform hexagons with complex domain wall
structures found in the experiments \cite{weston2019atomic,
rosen2020twist} are closely connected to the intrinsic lattice
reconstruction. The ``breathing'' of hexagons in a hexagonal
network of domain walls can give rise to distorted hexagons and
carry large entropy \cite{riste2012ordering}. Our calculations
suggest that these effects are realized for a general class of
domain wall networks in moir\'{e} materials (Reuleaux triangle to
hexagons). 

In Fig.~4a-b we compare the electronic band
structures of $\mathrm{tBLMoS_{2}}$ obtained for
$58.47^{\circ}$, which contains 24966 atoms. The lattice
reconstruction leads to an increment in the band-gap by
$\sim 20$ meV and significant changes in the spacing of
energy levels near the band edges. Interestingly, we find
the bands are ultra-flat (bandwidth $\lessapprox 1$ meV
within DFT) near the band-edges for both the structures
obtained with SR and SA. However, the wave-function
localizations corresponding to these flat bands are
strikingly different. To illustrate this, we show the
density associated with the first few-bands near the band
edges. With SR, the states near the valence band maximum
(VBM) resemble the states of a particle confined in a two
dimensional equilateral triangular well and are localized on
$\mathrm{AA^\prime}$ (Fig.~4c,4e) \cite{naik2019origin}. In
the reconstructed lattice, the degeneracies associated with
the equilateral triangular well are lifted, as triangles of
various shapes and depths are realized. In particular, the
wave functions corresponding to first three bands near the
VBM are localized on the $\mathrm{AA^\prime}$ stacking
enclosed by the trimer (Fig.~4h-j), whereas for the fourth
band, the wave function is localized on the larger
$\mathrm{AA^\prime}$ stackings (Fig.~4k). Since the area
enclosed by the trimer in the reconstructed lattice
(Fig.~4d) is $\theta$ independent, the spatial extension of
the localized VBM is expected to be $\theta$ independent.
The states near the conduction band minimum (CBM) are
localized on the $\mathrm{AB^\prime}$ stacking (Fig.~4f,
4l), whose size is also $\theta$ independent. This explains
the experimentally observed large tunnelling current at
$\mathrm{AB^\prime}$ \cite{weston2019atomic,
rosen2020twist}. The distinct spatial localizations of
electrons and holes originate from an in-plane strain driven
moir\'{e} potential \cite{naik2019origin}. For the
un-reconstructed lattice, the height of the moir\'{e}
potential is identical at all $\mathrm{AA^\prime}$ stackings
($\approx 132$ meV). After lattice reconstruction, the height
of the moir\'{e} potential at $\mathrm{AA^{\prime}}$
enclosed by the trimer is maximal ($\approx 171$ meV) and at
other $\mathrm{AA^{\prime}}$ stackings is unmistakably
smaller ($\approx 98$ meV) (SI, Sec. V for details). The
depth of the moir\'{e} potential at $\mathrm{AB^\prime}$
also changes after lattice reconstruction ($\approx - 306$
with SR and $\approx -255$ with SA, in meV). Furthermore, we have also obtained fully relaxed structures with SR and SA using DFT
calculations for a $\sqrt{3}\times \sqrt{3}\times1$
supercell of 58.53$^{\circ}$ tBLMoS$_2$. After applying a
small ($s>1.5\%$ )compressive strain, we show that the
lattice reconstructed structures obtained from SA are more
stable (SI, Sec. VI).

We have demonstrated reconstruction of the moir\'{e}
lattices of TMDs for $\theta>58.5^{\circ}$. These structures
can be probed using electron microscopy, optical imaging,
etc., and are expected to be generic for tBLs with different
sub-lattice atoms, including TMD heterostructures
\cite{alden2013strain,yoo2019atomic,weston2019atomic,rosen2020twist,zhang2018strain,xu2020nature,anderson2019moire,scuri2020electrically,baek2020high,ni2019soliton,
jiang2016soliton,holler2020low,gadelha2020lattice}.

\begin{acknowledgments}
We thank the Supercomputer Education and Research Centre at
IISc for providing computational resources, and Mit Naik,
Shinjan Mandal, Sudipta Kundu for several useful
discussions. H.~R.~K. thanks the Science and Engineering
Research Board of the Department of Science and Technology,
India for support under grant No. SB/DF/005/2017.
\end{acknowledgments} 

\setcounter{figure}{0}
\renewcommand{\thefigure}{S\arabic{figure}}

\title{Supplementary Information: \\ Reconstruction of moir\'{e} lattices in twisted transition metal dichalcogenide bilayers}

\author{Indrajit Maity, Prabal K. Maiti, H.~R.~Krishnamurthy}

\author{Manish Jain}
\email{mjain@iisc.ac.in}

\affiliation{
	Centre for Condensed Matter Theory, Department of Physics, \\ Indian Institute of Science, Bangalore-560012
}%

\maketitle

\newpage 
\section{I: Simulation details}
\textbf{Classical force-field based calculations:} \\
\textit{\textbf{(A) Details of structural relaxation}:} The standard relaxation (SR) is performed with the target pressure of P = 0 bar. For the simulated annealing (SA), the simulation box dimensions were kept the same as in SR. We use Nos\'{e}-Hoover thermostat while performing simulations with canonical ensemble. It should be noted that annealing at higher temperatures ($>1$ K) produces results similar to the ones discussed in the main text. In total, we have simulated $\sim 70$ twist angles ($\sim 40$ of them near $60^{\circ}$ and $\sim 30$ of them near $0^{\circ}$ ), with systems containing $10^5-10^7$ atoms, within the twist angle range, $0.2^{\circ} \leq \theta \leq 59.8^{\circ}$. For total energy comparisons as in Fig.3 of main text we use energy tolerance of $10^{-11}$ to perform the relaxation.  

\textit{\textbf{(B) Details of phonon calculations}:} While computing the phonon frequencies we have used force tolerance of $10^{-6}$ eV/\AA\ for the results presented in the main text. The relaxed structures are obtained by SR method.

\textit{\textbf{(C) Radial distribution function}:} The radial distribution function at the moir\'{e} scale, $g_{\mathrm{m}}(r)$ is computed with $9\times9\times1$ moir\'{e} supercell and averaged over 30 configurations. After performing molecular dynamics on a $3\times3\times1$ supercell, we replicate the structure to further create $9\times9\times1$ supercell and perform molecular dynamics. Identifying the moir\'{e} lattice points (MLPs) corresponding to $\mathrm{AA/ A^\prime B}$ from the interlayer separation (ILS) landscape, we calculate the $g_{\mathrm{m}}(r)$ defined as, $\frac{\langle N(r+ \delta r) \rangle}{A(\delta r)}$ with $\langle N(r+ \delta r) \rangle$ representing the average number of MLPs within a ring of radius $r$, width $\delta r$ and the area $A(\delta r)$. 

\textbf{Quantum simulations: } We use a double-$\zeta$ plus polarization basis for the expansion of wavefunctions. For all the electronic structure calculations we use the $\Gamma$ point in the moir\'{e} Brillouin zone to obtain the converged ground state charge density. A large vacuum spacing of $\sim 40$ \AA\ is used in the out-of-plane direction for all the density functional theory (DFT) calculations. We use a plane wave energy cut-off of 80 Ry to generate the 3D grid for the simulation. To check the adequacy of the effects of the small energy cut-off, we also simulate a $\mathrm{tBLMoS_2}$ with large twist angle ($50.6^{\circ}$, 222 atoms) using this cutoff as well as a larger cut-off of 320 Ry. We find a negligibly small difference in the electronic band structures (obtained with the two different cut-offs 320 Ry and 80 Ry). We do not include spin-orbit coupling in our DFT calculations. To study the effects of lattice reconstruction on electronic structure of the twisted bilayer $\mathrm{MoS_{2}}$ we simulate a $\sqrt{3}\times\sqrt{3}\times1$ moir\'{e} supercell for $\theta = 58.47^{\circ}$, which contains 24966 atoms. On this supercell, we perform SR (which does not show lattice reconstruction i.e. moir\'{e} periodicity of unrelaxed structure is preserved) and SA (which shows lattice reconstruction leading to trimerization i.e. moir\'{e} periodicity of un-relaxed structure is not preserved), as described in the main text. We perform electronic structure calculations in two ways: 

\textit{\textbf{(D) Multiscale simulations}:} We use the relaxed structures obtained with accurate classical force-field based simulations and carry out DFT calculations. In this approach, we do not further relax the structures, and local density approximation is used for the exchange-correlation functional. We employ the norm-conserving pseudopotentials \cite{troullier1991efficient}. This approach critically depends on the accuracy of the classical force-fields. The interlayer classical forcefield parameters were obtained by fitting the interlayer binding energy landscape obtained from the van der Waals corrected DFT calculations \cite{Mit_kc_2018}. This multiscale approach is widely used in predicting the electronic properties of the moi\'{e} materials \cite{Mit_kc_2018, naik2019origin}. All the electronic structure calculations reported in the main text (Fig.~4, for example) are obtained with this approach unless otherwise specified.

\textit{\textbf{(E) Relaxation using DFT}:} We use the relaxed structures obtained with classical simulations as a starting configuration and further perform relaxation with DFT. In this approach, we use van der Waals density functional that includes a non-local energy functional \cite{dion2004van} alongside Cooper exchange \cite{cooper2010van} as implemented in SIESTA. We employ norm-conserving, scalar relativistic pseudopotential obtained from PseudoDojo~\cite{van2018pseudo}. The relaxations with DFT are performed with 100 meV/\AA\ as the maximum atomic force tolerance for any atom. It should be noted that we use a relatively large atomic force tolerance as the system under consideration contains 24966 atoms. However, we find only a small fraction of atoms ($\approx 0.6\%$) has atomic forces greater than 40 meV/\AA\ in the fully relaxed structure. The atomic force tolerance of 40 meV/\AA\ is the default in SIESTA. Therefore, we expect the ordering of total energetics as shown in table~\ref{energetics} to be reliable. All the related structural relaxation calculations are performed with 4800 cores on a CRAY XC40 machine over 100 days' worth of computing (total: 4,80,000 core-hours) using SIESTA.

We obtain the ground state charge density using convergence criteria on both densities (with a tolerance of $10^{-4}$) and Hamiltonian (with a tolerance of $10^{-3}$ eV). We use the recently developed Pole Expansion, and Selected Inversion (PEXSI) technique \cite{lin2009fast,yu2018elsi,lin2014siesta} to construct the ground state charge density of large systems. This enables us to perform full DFT relaxation and compute the electronic properties of large-scale moir\'{e} patterns of TMDs within a reasonable time. We use the obtained ground state charge density and diagonalize to compute the electronic band-structure. We have also computed the ground-state charge density using standard diagonalization techniques and computed the electronic structure of $\sqrt{3}\times\sqrt{3}\times1$ lattice reconstructed structures. We find negligibly small differences in the electronic properties between these two approaches.

\clearpage 
\newpage  
 
\section{II : Reconstruction of moir\'{e} lattices for other TMDs}

\begin{figure}[b]
\includegraphics[scale=0.19]{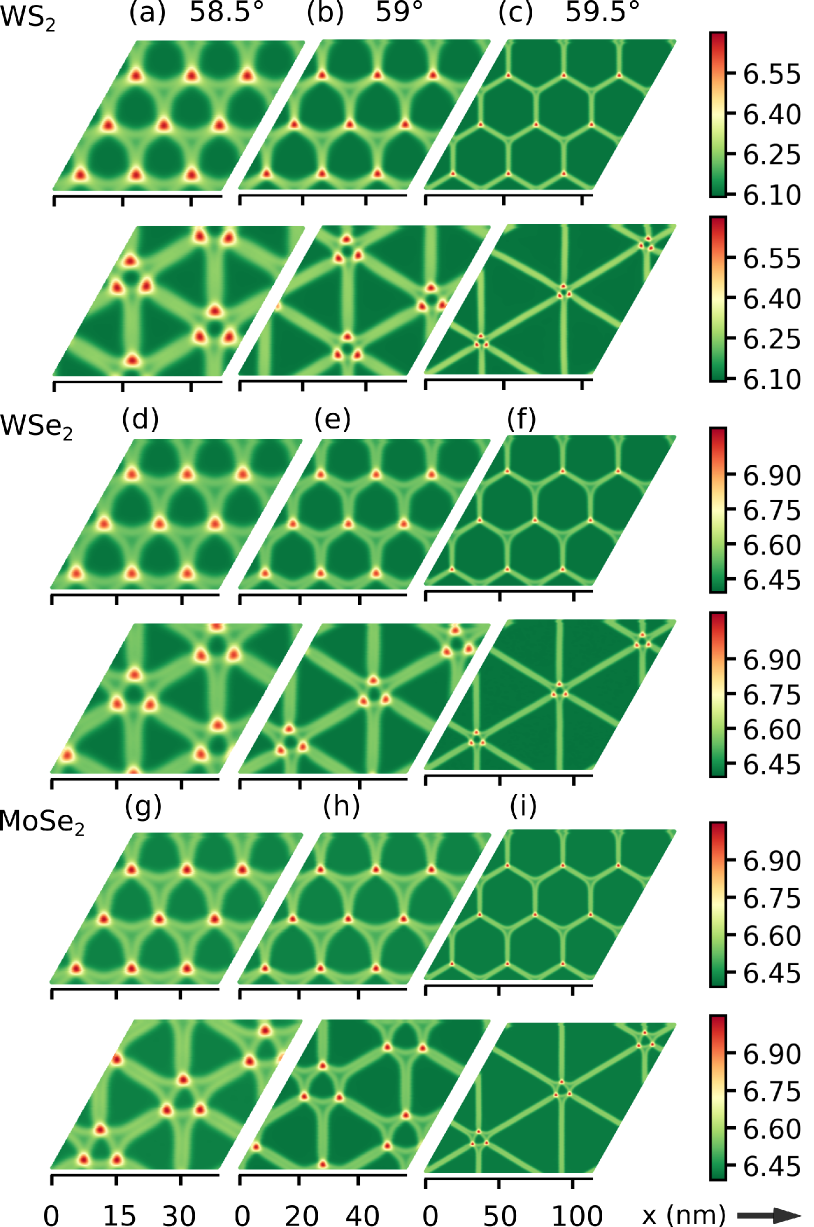}	
\caption{Interlayer separation landscape for for $\mathrm{WS_2\, WSe_2, \ MoSe_2}$ using standard relaxation (upper panel for each material) and simulated annealing (lower panel for each material) computed with a $3\times 3\times 1$ moir\'{e} supercell. The scales of the associated colorbars are for the interlayer separation in \AA.} 
\end{figure}

\newpage 
\section{III : Estimation of domain wall length}
The estimation of the domain wall length, $l$ is done in four steps : (i) Identifying approximately the regions exclusively belonging to the domain walls (ii) Defining a polygon that encloses all the points for a representative domain wall (iii) Finding a suitable representation of the \textit{thick} domain wall as a line. (iv) Averaging over the supercell to get statistically significant results. We illustrate this in Fig.~S2.
\begin{figure*}[!htbp]
\includegraphics[width=\textwidth]{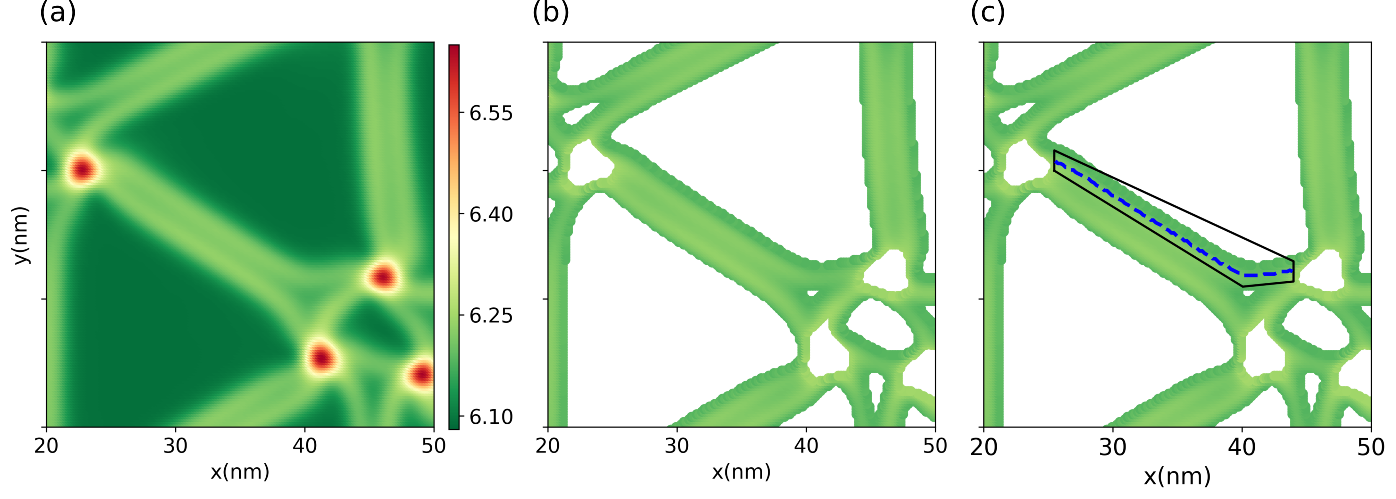}
\caption{ (a) Interlayer separation landscape using the simulated annealing approach for $\theta=59^{\circ}$. The scales of the associated colorbar are for the interlayer separation in \AA. (b) Identification of the domain wall region from ILS landscape (c) Defining the polygon (solid black line) and finding a suitable representation of the domain wall (dashed blue line).} 
\end{figure*}

\newpage 

\section{IV : Transient structures as $\theta\to60^{\circ}$}
\textbf{With $3\times 3\times 1$ moir\'{e} supercell}
\begin{figure*}[!ht]
\includegraphics[width=0.83\textwidth]{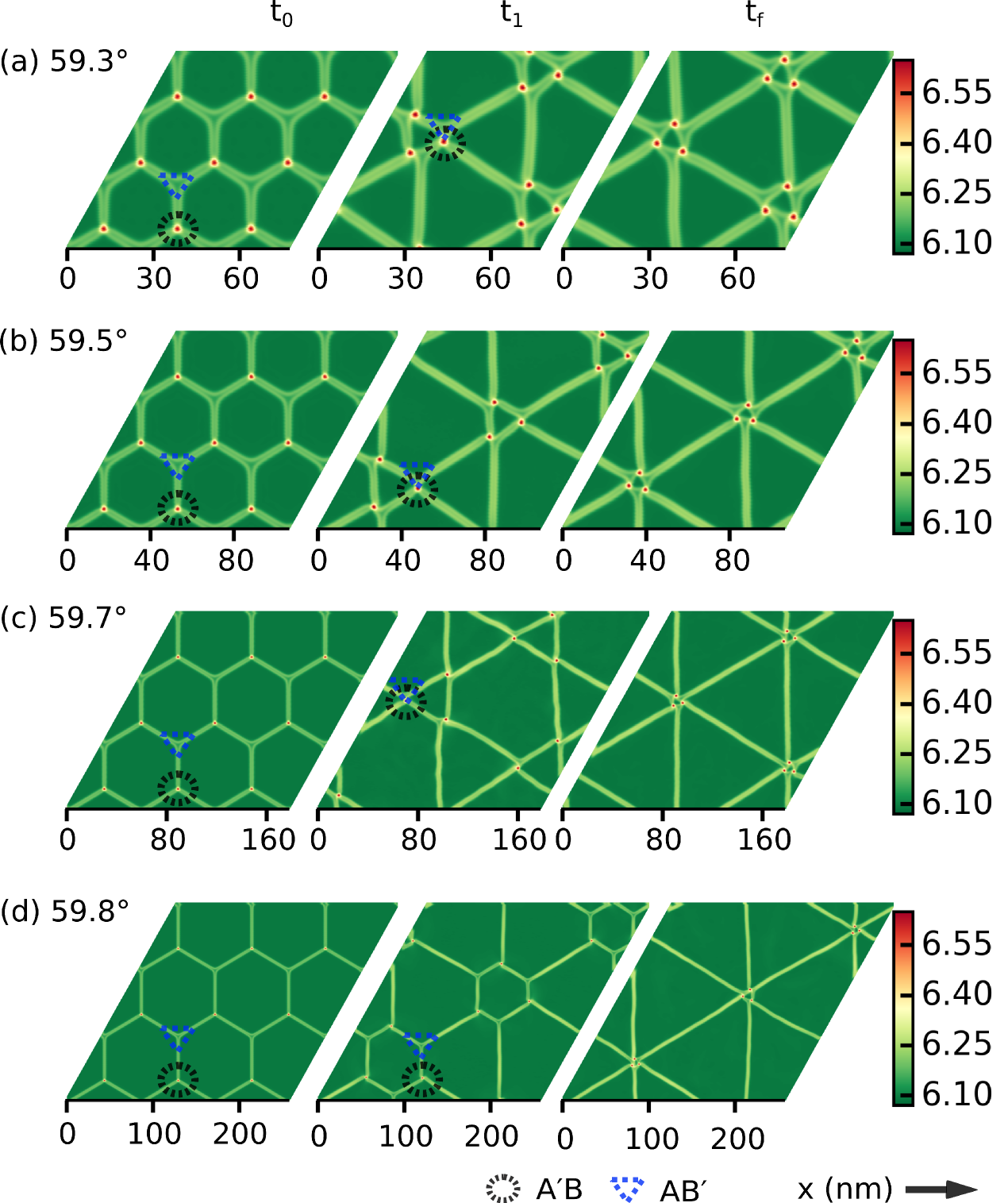}
\caption{Interlayer separation landscape using the simulated annealing approach during the equilibration of molecular dynamics simulations of a $3\times 3\times 1$ supercell of $\mathrm{tBL\ MoS_{2}}$. $t_0$ denotes results with standard relaxation, $t_1$ labels results before completely equilibrating, $t_f$ denotes results when the trimerization is complete (most stable). Similar results are obtained for other TMDs, as well. The scales of the associated colorbars are for the interlayer separation in \AA. Only $\mathrm{A^\prime B ,\ AB^\prime}$ stackings are marked for a few cases. }
\end{figure*}
\newpage 
\textbf{With $10\times 10\times 1$ moir\'{e} supercell}

Furthermore, we have also simulated a moir\'{e} supercell containing 100 moir\'{e} unit cells ($10\times 10\times 1$). This helps us in establishing two important aspects: 
 
(a) The larger number of degrees of freedom can result in complicated lattice reconstructed structures. Some of the transient structures found here resemble the structures observed in recent experiments \cite{weston2019atomic, rosen2020twist}. Our results suggest that the distorted hexagons observed in the experiment are results of intrinsic lattice reconstruction. Although, the external substrate, strain etc. can make the transient structures found in our calculations metastable or stable.
 
(b) Since the $10\times10\times1$ supercell contains 100 $\mathrm{A^\prime B}$ stackings, a complete trimerization can not be formed (as 100/3 is not an integer). Therefore, complex reconstruction of moir\'{e} lattices (other than trimerization) is realized. 

\begin{figure*}[!htbp]
	\includegraphics[width=\textwidth]{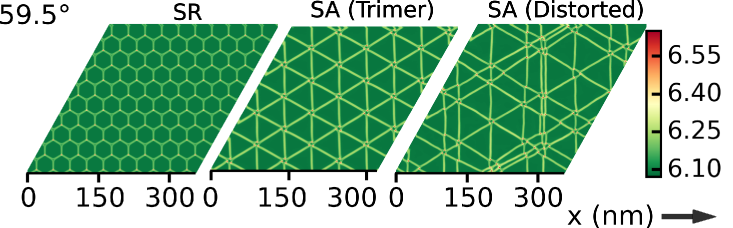}
	\caption{(a): Interlayer separation landscape of $\mathrm{tBLMoS_{2}}$ of lattice un-reconstructed structures, (b): lattice reconstructed structures consisted of only trimers, and (c): lattice reconstructed structures with large parallel domain walls. The scales of the colorbar, in \AA\ and corresponds to interlayer separation.}
	\label{entropy}
\end{figure*}

We find that the simplest example of lattice reconstruction of moir\'{e} lattice is trimerization of the unfavourable stackings. The corresponding figures are shown in the main text (see Fig.~1). However, we find that other complicated lattice reconstructed structures are also possible. It should be noted that the energy difference between the trimerized and other lattice reconstructed structures (such as distorted hexagons, etc.) are only a few tens of meV per moir\'{e} lattice. Therefore, the difference between these structures is primarily entropic. To illustrate this, we compute the total energies of the un-reconstructed ($E^{\mathrm{SR}}$), lattice reconstructed structures containing only trimers ($E^{\mathrm{SA}}_{trimer}$), and lattice reconstructed structures containing kagome-like patterns, large parallel domain walls, ($E^{\mathrm{SA}}_{distorted}$) see table~\ref{tab:entropic}. It is evident from the table that both the lattice reconstructed structures obtained with SA are significantly energetically lower than the lattice un-reconstructed structures obtained with SR. The difference in energy is $\approx -7$ eV per moir\'{e} unit-cell. However, the difference between the energies of the different lattice reconstructed structures is $\approx 50$ meV per moir\'{e} unit-cell. For clarity, all the three structures are shown in Fig.~\ref{entropy}. We expect that the distorted reconstructed structures  such as the one shown in Fig. S4 (c) have higher entropy, and might therefore become the favored structures at finite temperatures via entropic stabilization. 

\begin{table}[h]
	\begin{tabular}{|m{3cm}|m{3cm}|m{3cm}|m{3cm}|}
		\hline
		Twist angle & $E^{\mathrm{SR}}$ & $E^{\mathrm{SA}}_{trimer}$ & $E^{\mathrm{SA}}_{distorted}$ \\ 
		\hline
		59.5$^{\circ}$ & -98699.294 eV & -98706.667 eV & -98706.611 eV \\
		\hline 
	\end{tabular}
	\caption{Total energies per moir\'{e} lattice computed used force-field based simulations. $E^{\mathrm{SR}}$ and $E^{\mathrm{SA}}_{distorted}$ are computed with a moir\'{e} supercell containing 100 moir\'{e} lattices ($10\times 10\times 1$) and $E^{\mathrm{SA}}_{trimer}$ is computed with a moir\'{e} supercell containing 81 moir\'{e} lattices ($9\times9\times1$).}
	\label{tab:entropic}
\end{table} 

\begin{figure*}[!ht]
\includegraphics[width=0.9\textwidth]{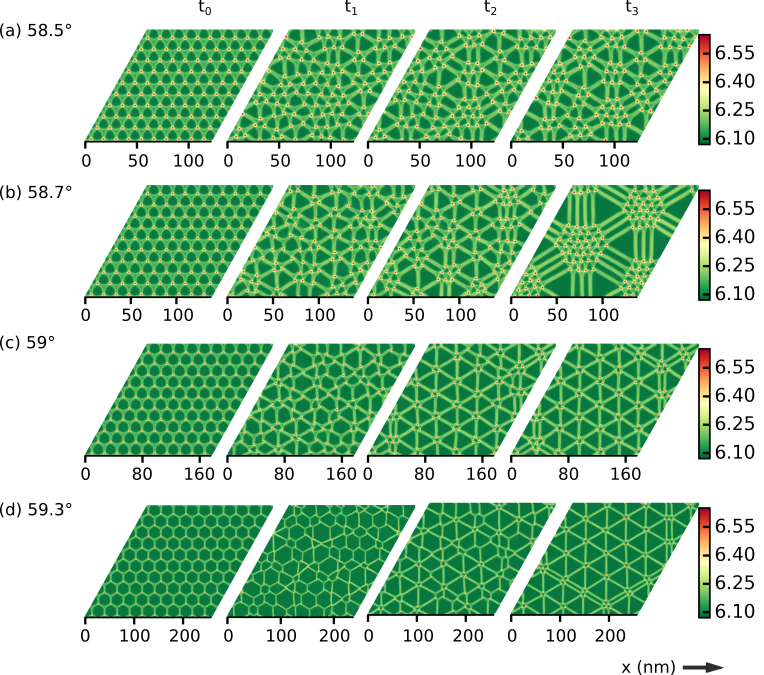}	
\caption{ (a) Interlayer separation (ILS) landscape using the simulated annealing approach for several twist angles with $10\times 10 \times 1$ moir\'{e} supercells. The scales of the associated colorbars are for the interlayer separation in \AA. For every twist angle four ILS landscapes are shown, where the snapshots for annealing are taken from 1-1.5 nanosecond molecular dynamics runs. $t_0=0$ ns, $t_1\sim 0.1-0.2$ ns, $t_2\sim 0.3-0.4$ ns,  and $t_3\sim1-1.5$ ns i.e. $t_0<t_1<t_2<t_3$. }
\label{sa_10x10x1_1} 
\end{figure*}

\begin{figure*}[!ht]
\includegraphics[width=0.85\textwidth]{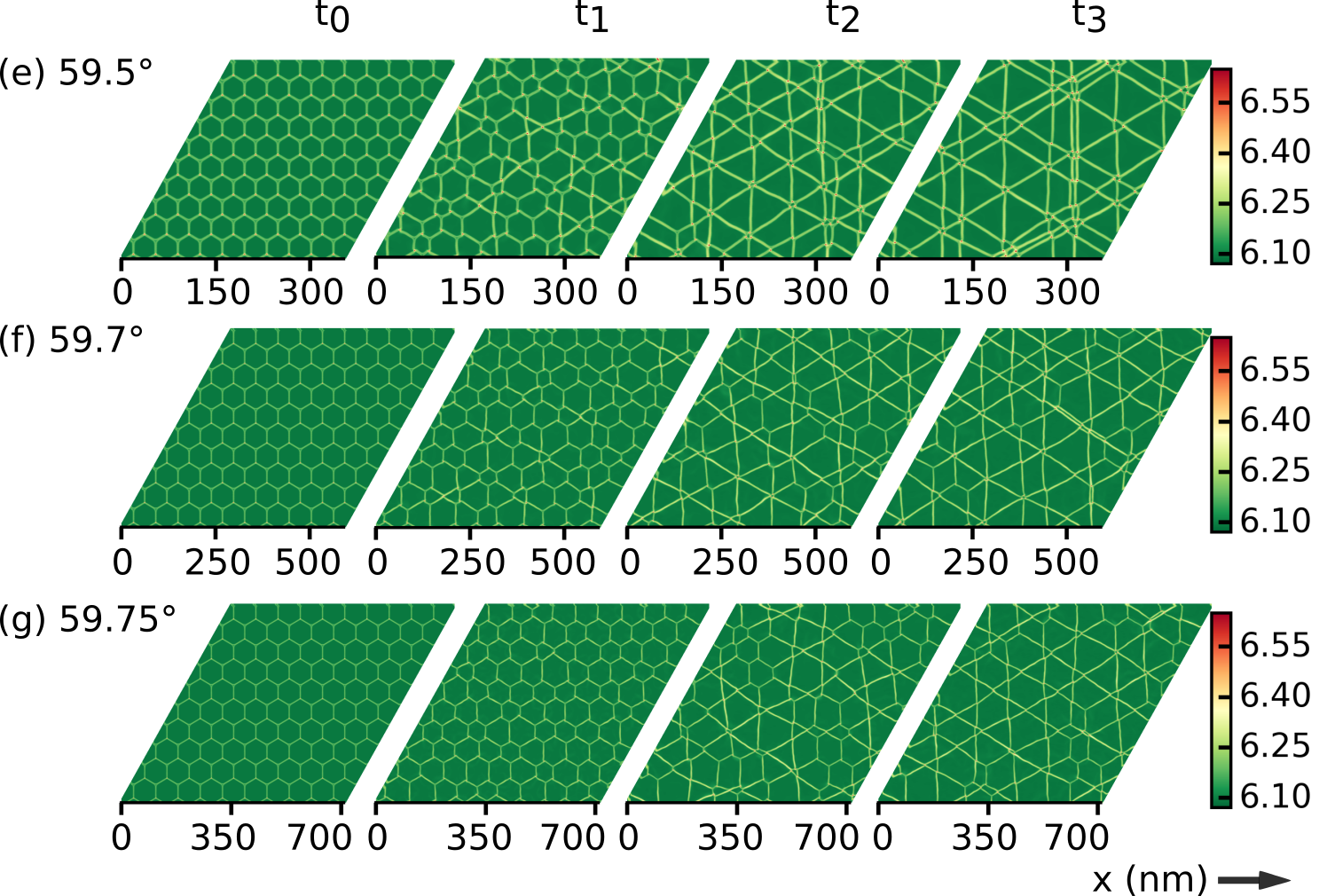}	
\caption{ (a) Interlayer separation (ILS) landscape using the simulated annealing approach for several additional twist angles with $10\times 10 \times 1$ moir\'{e} supercells. The scales of the associated colorbars are for the interlayer separation in \AA. For every twist angle four ILS landscapes are shown, where the snapshots for annealing are taken from 1 nanosecond molecular dynamics runs. $t_0=0$ ns, $t_1\sim 0.3-0.4$ ns, $t_2\sim 0.6$ ns and $t_3\sim1$ ns i.e. $t_0<t_1<t_2<t_3$. }
\label{sa_10x10x1_2} 
\end{figure*}
\clearpage 

\newpage 

\section{V : Confining potential from DFT}
\begin{figure*}[!ht]
\includegraphics[width=0.9\textwidth]{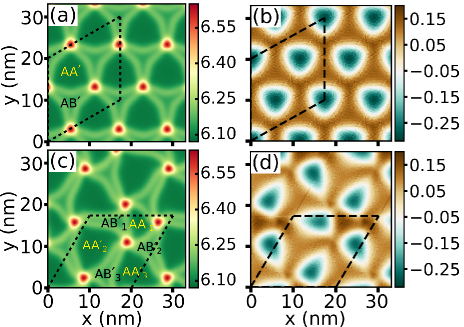}	
\caption{(a), (b): Interlayer separation landscape and the confining moir\'{e} potential, respectively, for structures obtained with SR. (c), (d) Interlayer separation landscape and the confining moir\'{e} potential, respectively, for structures obtained with SA. The scales of the associated colorbars for the interlayer separation are in \AA. The scales of the associated colorbars for moir\'{e} potential are in eV. The moir\'{e} $\sqrt{3}\times\sqrt{3}\times1$ cell is marked with black dashed lines. Different stacking locations are marked in (a), (b).}
\label{pot} 
\end{figure*}

To compute the effective moir\'{e} potential, we use the following steps:- (a) We average the self-consistent total DFT potential in the out-of-plane direction. (b) We create a map of the local potential using the Voronoi cells obtained from the positions of the Mo atoms ($x_{\mathrm{Mo}},y_{\mathrm{Mo}}$) of bottom layer $\mathrm{MoS_{2}}$ and obtain the macroscopic potential ($V_{\mathrm{M}}(x_{\mathrm{Mo}},y_{\mathrm{Mo}})$). (c) We compute the confining potential by subtracting the average: $V_{\mathrm{conf}}=V_{\mathrm{M}}-\bar{V}_{\mathrm{M}}$. The confining potential is plotted in Fig~\ref{pot}. The electrons (holes) are localized at the $\mathrm{A^\prime B}$ ($\mathrm{AA^\prime}$) stacking where the confining potential has a minimum (maximum). The minimum and maximum of the potential at these stackings are tabulated in table~\ref{pot_moire}. All these calculations are performed using multi-scale simulations, as outlined above. We also tabulate the first few eigenvalues near VBM and CBM for both un-reconstructed structures (table~\ref{bands_sr}) and lattice reconstructed structures (table~\ref{bands_sa}). As can be seen from the table, the separation between the first few bands near the band edges and associated degeneracies change significantly.  

\begin{table}[h]
\begin{tabular}{|m{3cm}|m{2.2cm}|m{2.2cm}|m{2.2cm}|m{2cm}|m{2cm}|m{2cm}|}
\hline
Structure & $V_{\mathrm{AA^\prime_{1}}}$ (meV) & $V_{\mathrm{AA^\prime_{2}}}$ (meV) & $V_{\mathrm{AA^\prime_{3}}}$ (meV) & $V_{\mathrm{AB^\prime_{1}}}$ (meV) & $V_{\mathrm{AB^\prime_{2}}}$ (meV) & $V_{\mathrm{AB^\prime_{3}}}$ (meV) \\ 
\hline
SR ($\sqrt{3}\times\sqrt{3}\times1$) & 132  & 132 & 132 & -306 & -306 & -306 \\
\hline 
SA ($\sqrt{3}\times\sqrt{3}\times1$) & 171  & 97 & 103 & -240 & -268 & -258 \\
\hline 
\end{tabular}
\caption{Moir\'{e} potential height at different stackings for the unreconstructed structures obtained with SR and the lattice reconstructed structures obtained with SA. }
\label{pot_moire}
\end{table}

\begin{table}[h]
\begin{tabular}{|m{2.8cm}|m{1.8cm}|m{1.8cm}|m{1.8cm}|m{1.8cm}|m{1.8cm}|m{1.8cm}|m{1.8cm}|}
\hline
& $\mathrm{V_{1}}$ (eV) & $\mathrm{V_2}$ (eV) & $\mathrm{V_3}$ (eV) & $\mathrm{V_4}$ (eV) & $\mathrm{C_1}$ (eV) & $\mathrm{C_2}$ (eV) & $\mathrm{C_3}$ (eV) \\ 
Structure & [d] & [d] & [d] & [d] & [d] & [d] & [d] \\  
\hline
SR ($\sqrt{3}\times\sqrt{3}\times1$) & -4.342  & -4.365 & -4.381 & -4.395 & -3.394 & -3.376 & -3.337 \\ 
 & [$1\times3$] & [2$\times$3] & [$1\times 3$] & [$2\times 3$] & [$6\times 3$] & [$2\times 3$] & [$2\times 3$] \\
\hline 
\end{tabular}
\caption{First few bands near the valence band maximum and the conduction band minimum with the degeneracies specified in brackets.}
\label{bands_sr}
\end{table}

\begin{table}[h]
\begin{tabular}{|m{2.8cm}|m{1.8cm}|m{1.8cm}|m{1.8cm}|m{1.8cm}|m{1.8cm}|m{1.8cm}|m{1.8cm}|}
\hline
& $\mathrm{V_{1}}$ (eV) & $\mathrm{V_2}$ (eV) & $\mathrm{V_3}$ (eV) & $\mathrm{V_4}$ (eV) & $\mathrm{C_1}$ (eV) & $\mathrm{C_2}$ (eV) & $\mathrm{C_3}$ (eV) \\ 
Structure & [d] & [d] & [d] & [d] & [d] & [d] & [d] \\  
\hline
SA ($\sqrt{3}\times\sqrt{3}\times1$) & -4.32 & -4.365 & -4.372 & -4.379 & -3.354 & -3.352 & -3.348 \\ 
 & [1] & [1] & [2] & [1] & [2] & [3] & [4] \\
\hline 
\end{tabular}
\caption{First few bands near valence and conduction bands with the degeneracies specified in brackets.}
\label{bands_sa}
\end{table}

\clearpage

\section{VI : Structural relaxation results using DFT}
The unit-cell lattice constant of the $\mathrm{BLMoS_2}$ is 3.138 \AA and the optimum interlayer separation for the most stable $\mathrm{AA^{\prime}}$ stacking is 6.1 \AA. Using classical force-field based simulations, we find the reconstruction of the moir\'{e} lattices occur for $\theta>58.5^{\circ}$. The characteristic angle of $\theta_{c}\approx 58.5^{\circ}$ is dependent on both the intralayer elastic energy parameters and the stacking energy parameters. Therefore, we expect the characteristic angle for the onset of lattice reconstruction with DFT to be slightly different from that seen in force-field-based simulations. To prove purely using DFT that the reconstructed structures obtained with classical force-field based simulations are more stable than the un-reconstructed structures, one would have to perform several huge DFT calculations. This is, of course, \textit{practically impossible} as the system contains $2.5\times 10^4$ to $10^6$ atoms. One possible strategy to circumvent this immensely computationally expensive task is to tune the characteristic angle for the onset of lattice reconstruction and reduce the moir\'{e} supercell size on which DFT based relaxations are performed. This can be achieved by applying a small compressive strain.

\begin{figure*}[!ht]
	\includegraphics[width=0.8\textwidth]{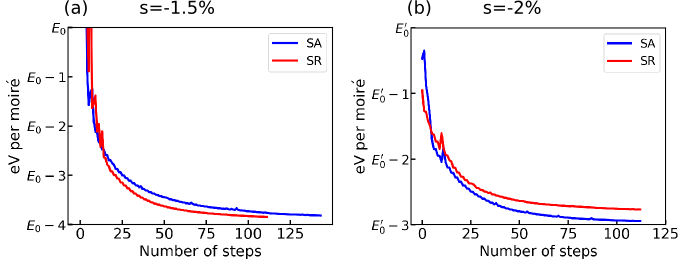}	
	\caption{The change in total energies during the relaxation in SIESTA for a compressive strain of $-1.5\%$ (in (a)) and $-2\%$ (in (b)). $E_{0}$ (=-6968752 eV) and $E_{0}^\prime$ (=-6968695 eV) are the offsets for the total energy. SR denotes standard relaxation and SA denotes simulated annealing.}
	\label{dft_rel} 
\end{figure*}

We have applied a series of small compressive strains and created the moir\'{e} lattices of $\mathrm{MoS_{2}}$ for $\theta = 58.53^{\circ}$ with systems containing 24966 atoms. We have performed SR and SA on the $\sqrt{3}\times\sqrt{3}\times1$ supercell of the moir\'{e} lattice using force-field based simulations. Using these structures as starting configurations, we have performed structural relaxation using DFT with SIESTA. As can be seen from the table~\ref{energetics}, the lattice reconstructed structures obtained with SA becomes more stable than the lattice un-reconstructed structures obtained with SR when the compressive compressive strain exceeds a threshold. Thus, we have established the reconstruction of moir\'{e} lattices in twisted TMD bilayers close to $60^{\circ}$ twist angles, using both classical force-field based calculations and first-principles based DFT calculations. We show the change in total energies per moir\'{e} unit-cell during the relaxation based on DFT in Fig~\ref{dft_rel}.

\begin{table}[h]
	\begin{tabular}{|m{4cm}|m{3cm}|m{3cm}|m{3cm}|}
		\hline
		$\theta$ (supercell) & Strain & Moir\'{e} supercell & $\Delta E$ per moir\'{e} \\
		&  & $\sqrt{3}a_{\mathrm{M}}$ &  \\ 
		\hline
		58.47$^{\circ}$ ($\sqrt{3}\times\sqrt{3}\times1$) & 0\%  & 202.42 \AA & 8.4 eV \\
		\hline 
		58.47$^{\circ}$ ($\sqrt{3}\times\sqrt{3}\times1$) & -0.7\% & 201 \AA & 1.8 eV \\
		\hline 
		58.47$^{\circ}$ ($\sqrt{3}\times\sqrt{3}\times1$) & -1.5\% & 199.3 \AA & 0.03 eV \\
		\hline 
		58.47$^{\circ}$ ($\sqrt{3}\times\sqrt{3}\times1$) & -2 \% & 198.03 \AA & -0.175 eV   \\
		\hline 
	\end{tabular}
	\caption{$\Delta E = (E^{\mathrm{SA}}-E^{\mathrm{SR}})/3$: Total energy differences between the trimerized structure (SA) and lattice un-reconstructed structure (SR) for several small compressive strains computed after relaxations performed with DFT.}
	\label{energetics}
\end{table}  

\begin{figure*}[!ht]
	\includegraphics[width=0.5\textwidth]{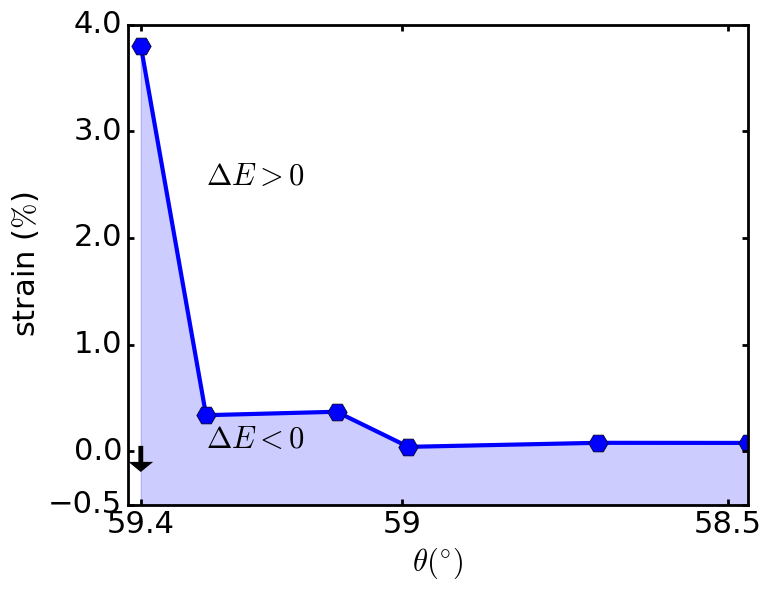}	
	\caption{Strain-twist angle phase diagram computed with force-field based simulations. When the energy difference between lattice reconstructed and un-reconstructed structures ($\Delta E$) is negative, the lattice reconstructed structures are more stable (the shaded region). Under the application of large bi-axial tensile strain, the lattice un-reconstructed structures obtained with SR become energetically more stable. All the calculations are performed on a $\sqrt{3}\times\sqrt{3}\times1$ supercell for several twist angles with classical force-field based simulations.}
	\label{figS9} 
\end{figure*}

We also compute the bi-axial strain dependence of the energetics between structures obtained with SR and SA using force-field based simulations. We find the energy differences between these structures are sensitive to strain (Fig.~\ref{figS9}). The lattice reconstructed structures are always more stable than the structures obtained with SR for any biaxial compressive strain. However, under sufficient tensile strain, the structures obtained with SR are more stable than the lattice reconstructed structures. The largest tensile strain applied in our simulation is $\approx 3.8 \%$

\newpage


%

\end{document}